# Mechanochemical Nano-Writing of an Atomically Thin Metal

Shuai Zhang,[1,3†] Yanyu Jia,[2†] Atanu Samanta,[4] Yutian Bao,[4] Haosen Guan,[2] Zhaoyi Joy Zheng,[2,5] Guangming Cheng,[6] Ting Liu,[7] Cangyu Qu,[1] Kenji Watanabe,[8] Takashi Taniguchi,[9] Nan Yao,[6] Ashlie Martini[7], Leslie Schoop,[10] Andrew M. Rappe,[4] Sanfeng Wu,[2]* Robert W. Carpick[1]*

[1]Dept. of Mechanical Engineering and Applied Mechanics, University of Pennsylvania, Philadelphia, PA, 19104, USA
[2]Dept. of Physics, Princeton University, Princeton, NJ, 08544, USA
[3]Center for X-mechanics, Dept. of Engineering Mechanics, Zhejiang University, Hangzhou, 310027, China
[4]Dept. of Chemistry, University of Pennsylvania, Philadelphia, PA, 19104-6323 USA
[5]Dept. of Electrical and Computer Engineering, Princeton University, Princeton, NJ, 08544, USA
[6]Princeton Materials Institute, Princeton University, Princeton, NJ, 08544, USA
[7]Dept. of Mechanical Engineering, University of California Merced, Merced, CA, 95343, USA
[8]Research Center for Electronic and Optical Materials, National Institute for Materials Science, 1-1 Namiki, Tsukuba 305-0044, Japan
[9]Research Center for Materials Nanoarchitectonics, National Institute for Materials Science, 1-1 Namiki, Tsukuba 305-0044, Japan
[10]Dept. of Chemistry, Princeton University, Princeton, NJ 08544, USA

[†]These authors contributed equally to this work.
*Corresponding author. E-mail: sanfengw@princeton.edu; carpick@seas.upenn.edu

**Mechanical energy accelerates many physicochemical processes, including materials syntheses that are hard to produce with thermal energy alone. However, physical understanding connecting applied mechanical forces with internal stresses and ensuing reaction mechanisms is lacking. Here we demonstrate mechanical force-enabled synthesis and nanoscale patterning to metallize a two-dimensional (2D) material, producing an atomically-thin superconducting material. Localized force applied by atomic force microscope tips to van der Waals (vdW) encapsulated stacks of 2D bilayer $MoTe_2$ and adjacent source Pd guides 2D $Pd_7MoTe_2$ growth with 50 nm lateral resolution. Force accelerates reaction kinetics exponentially per Eyring's stress-assisted thermal activation model, reducing synthesis temperatures from ~200°C to near-room temperature. Finite element simulations, density functional theory, and ab-initio grand canonical Monte Carlo calculations show that tip-induced compression facilitates Pd chemisorption to tensile-strained $MoTe_2$ that converts to uniform $Pd_7MoTe_2$. This demonstrates a new,**

**generalizable paradigm for nanoscale synthesis of quantum materials, and high-precision engineering of superconductivity.**

## I. INTRODUCTION

Transformative progress in 2D materials have catalyzed efforts to create atomically-thin structures with broad ranges of functional properties and to integrate them into devices with superb capabilities and performances. This includes a need for patternable, atomically thin metals and superconductors which are highly desired for constructing advanced electronic devices. However, unlike layered van der Waals (vdW) crystals, metals are not mechanically exfoliable, and many deposited metals dewet on substrates to form discontinuous islands when the thickness is reduced to a few nanometers. Synthesizing uniform 2D metals hence requires innovative processes. Achieving this, especially in a controlled fashion, would open exciting opportunities in nanoelectronics (*1, 2*), photonics (*3*), plasmonics (*4*), and next-generation quantum devices (*5-8*) harnessing designed material structures in the 2D limit.

Recent examples of depositing or fabricating 2D metals include the use of molecular beam epitaxy to grow 2D Sn, Pb, In, and other elemental metals (*9-11*) on specific substrates. Freestanding 2D Fe and Cr have been created in suspended pores, although the lateral size in these cases was limited to nanometer scale (*12, 13*). Monolayer gold has been synthesized using a wet-chemical etching method (*14*). Long-distance 2D mass transport upon thermal annealing has been found to occur at interfaces between various combinations of metals and 2D transition metal dichalcogenides (TMD), leading to uniform layers of Pd and Ni of a precise 7-atom thickness, chemisorbed to the modified TMD template (*15-17*). More recently, mechanical squeezing applied to molten metals successfully created 2D films of Bi, Ga, Sn, and In with Å-thickness between TMD layers (*18, 19*). These methods are generalizable to other combinations of metal and vdW crystals. The emerging list of atomically thin metals and new 2D-metallized compounds obtained experimentally is laying a foundation for exploring new science and exploiting these materials in many forefront applications (*1-8*).

However, these approaches for synthesizing atomically thin metals developed so far all lack control in the growth process, leaving open severe challenges in processing and integrating such fragile 2D metals into sophisticated material structures and devices. It is thus critical to realize controlled synthesis with nanoscale precision, especially for in-plane features, without sacrificing the quality of 2D metals and their interfaces with adjacent materials. Here we introduce and demonstrate a mechanochemical nano-synthesis route, distinct from all previous methods, that achieves unprecedented precision in controlled in-plane growth, and consequently enables arbitrarily-patternable nano-writing on demand of atomically thin metals hybridized with 2D materials, including air- and temperature-sensitive 2D materials.

Our approach employs the advent of nanoscale mechanochemistry, which achieves nano-synthesis and nanomanufacturing by using mechanical force applied via a sharp tip in an atomic force microscope (AFM) to kinetically drive controlled chemical synthesis processes (*20-27*). A defining feature of this approach is its ability to precisely regulate applied stresses thanks to the accurate and precise force control capability of AFM which can reach the piconewton (pN) level. This can ensure minimal material damage, preserving structural integrity and functionality of the produced material, while the small nanoscale area over which the force is applied, as defined by the tip-sample contact area, can permit stresses to reach levels high enough to drive desired chemical reactions or other material transformations (*20, 23, 25*). Moreover, stress fields in the material can extend well beyond the location of the applied force (*28*), facilitating chemical reactions at locations remote from the immediate tip-material contact location, *e.g.*, underneath a protection layer. Recent studies have demonstrated mechanochemistry's ability to create amorphous, polycrystalline, and nanocomposite structures through tribomechanical and tribochemical interactions driven by localized contact forces (*20, 21, 29*). Below we present our discovery that AFM-based mechanochemical nano-synthesis can be employed to precisely control the growth of an atomically thin, crystalline metal in the form of a new 2D-metal-based compound with nanometer scale lateral dimensions and geometries.

## II. RESULTS

### A. Mechanochemical Synthesis and Nano-Writing of 2D Palladium

**Figure 1A** shows a schematic of the experimental setup. The sample consists of exfoliated monolayer or bilayer 2H-$MoTe_2$ transferred onto pre-deposited Pd electrodes, typically about 27 nm thick, on a $SiO_2$ substrate. The sample is encapsulated with a thin layer of exfoliated hBN to minimize oxidation during the atomic force microscope (AFM) measurements (see Methods and Fig. S1). Recent work by some of the present authors showed that heating a thin layer of a 2D TMD that includes regions adhered to a localized metal source drives the long-range transport of 2D metal atoms across the 2D material at temperatures well below the melting point of the bulk metal (*15-17*). These migrating atoms form covalent/metallic bonds with the host TMD and modify its 2D lattice substantially, resulting in stable 2D metal-TMD compounds in the form of single crystals templated by the 2D material that extend outward from the original bulk metal source (*15-17*). A typical example is the interaction between Pd and $MoTe_2$, where heat induces Pd atoms to migrate from the source and subsequently bond with $MoTe_2$, precisely forming one 7-atom thick Pd crystalline film per $MoTe_2$ layer, *i.e.*, a new 2D metalized compound, $Pd_7MoTe_2$ (*15-17*). Consistent with these prior results, upon heating the sample (sample D1, Pd on bilayer $MoTe_2$) at 200 °C for 5 minutes, a noticeable darkening of the optical contrast in regions extending outward from the localized Pd source is observed. An example for a region in sample D1 is shown in **Fig. 1B**, left panel, indicating the migration of Pd atoms that convert the $MoTe_2$ layer to the new 2D $Pd_7MoTe_2$ layer, a chemical process occurring within the 2D plane, independent of the substrate and encapsulation layers. This change is further confirmed by AFM topography measurements, where the region surrounding the electrode exhibits a height increase of $1.55 \pm 0.18$ nm *vs.* the adjacent $MoTe_2$ (**Fig. 1B,** right panel). This height increase is consistent with the previously reported value (*17*). The growth of $Pd_7MoTe_2$ thus involves two processes: long-distance transport of Pd atoms from the bulk source to the growth front via preformed 2D Pd layer, and then the formation of new $Pd_7MoTe_2$ structure at the growth front (*15*). This 2D chemical reaction is ubiquitous, as it occurs for multiple combinations of metals and TMDs (*15*).

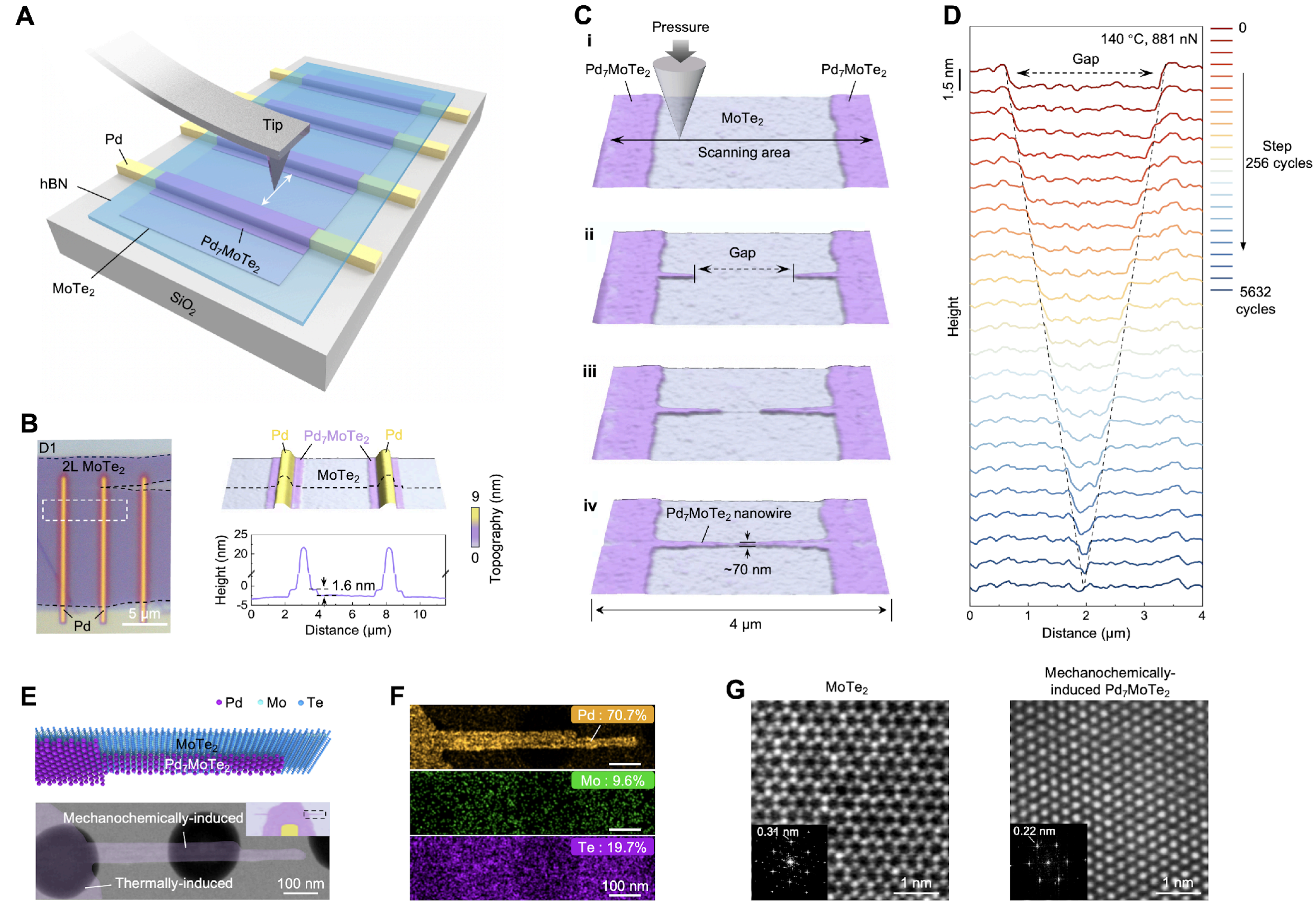

**FIG. 1. Mechanochemically induced diffusion of Pd and formation of $Pd_7MoTe_2$.** (**A**) A schematic of the experimental setup, where an AFM tip is scanning on the sample surface in contact mode. The stacking configuration of the device consists of a Pd strip electrode (≈27 nm thick) deposited on a silica substrate, overlaid with a bilayer of 2H-$MoTe_2$, and capped with a thicker hBN layer. (**B**) Typical optical microscope image of sample D1 (left panel); the AFM 3D topographic image is measured within the area marked by the white rectangle in the optical image (left panel), and the corresponding height profile is measured along the black dashed line (right panel). (**C**) The setup for mechanochemically induced growth of $Pd_7MoTe_2$ involves the AFM tip repeatedly scanning along a single line across $MoTe_2$ (central region) between two thermally grown $Pd_7MoTe_2$ regions (left and right purple regions) at 881 nN applied normal load; (i) to (iv) are four corresponding representative AFM topographic snapshots obtained after linear scanning for increasing numbers of cycles. (**D**) Representative selections of height profiles measured by the tip with increasing scan cycles. (**E**) A schematic of the sample structure (upper panel) and typical scanning transmission electron microscope (STEM) image of the $Pd_7MoTe_2$, comprising both the original thermally induced regions and the mechanochemically written nanowires (lower panel). Inset: AFM topographic image before the sample was transferred onto the TEM grid. (**F**) EDX analysis of $Pd_7MoTe_2$ nanowire. (**G**) Atomic-resolution STEM image of pristine $MoTe_2$ (in a region with no Pd) at left, and the mechanochemically induced $Pd_7MoTe_2$ at right. Insets are fast Fourier transforms of each image, revealing the characteristic lattice spacing. All experimental data are from sample D1.

We examine whether this reaction and 2D growth can be triggered by stress, instead of only thermal annealing. Mechanochemical synthesis has been demonstrated possible for other reactions (*20, 30, 31*), particularly at the nanoscale (*20*). To explore the effect of local stress, we applied localized mechanical force using an AFM probe (**Fig. 1C(i)**) in contact with the sample. This creates a stress field within the material, including at the buried Pd-$MoTe_2$ interface. We show that this stress field can indeed drive the reaction, promoting precisely controlled growth of 2D Pd on $MoTe_2$ and forming $Pd_7MoTe_2$.

In representative example, the AFM probe was brought into contact with the sample (D1) and repeatedly scanned at an applied normal load (force) of 881 nN along a line spanning a 2.7 μm-wide pristine $MoTe_2$ region (middle, colored in gray) in between two thermally grown $Pd_7MoTe_2$ regions (left and right, colored in purple), as marked by the black line. To minimize material growth driven solely by thermal effects and moreover to demonstrate that stress permits formation of the material at reduced thermal loads, the sample temperature was set at 140 °C, much lower than the previously reported thermally driven growth temperature of approximately 200 °C (*15*). Approximately every 2000 cycles, a zoomed-out topographic image was collected (**Fig. 1C**). Two nanowires progressively grew outward from the bilayer edges along the scanning line, and eventually merged. The nanowire height ≈1.6 nm is consistent with that of thermally grown $Pd_7MoTe_2$ (**Fig. 1D**); nanowire lengths of ≈520 nm to ≈1000 nm are seen after scanning 2,304 and 4,096 cycles respectively (**Fig. 1C(ii)** and **(iii)**), with the lines merging at after approximately 5,632 scans, forming a continuous nanowire with a minimum width of ≈70 nm (**Fig. 1C(iv)**). Both the topographic height and the interfacial friction force response (see Fig. S2) for the mechanochemically grown nanowires and the thermally grown $Pd_7MoTe_2$ are indistinguishable. This nanoscale mechanochemical synthesis was replicated with multiple tips, for samples with monolayer $MoTe_2$, and for different substrate materials (see Fig. S2, Fig. S3 and Fig. S4).

Sample D1 was then transferred onto a transmission electron microscopy (TEM) grid for structural and elemental characterization (see Methods). The inset of **Fig. 1E** presents a topographic image of another mechanochemically written nanowire before transfer, and the corresponding large-scale TEM image after transfer is shown in **Fig. 1E**. The structure remains intact post-transfer, preserving both the thermally and mechanochemically induced structures. Energy-dispersive X-ray spectroscopy (EDX) imaging (**Fig. 1F**) shows a homogeneous distribution of Pd, Mo, and Te within the nanowires. The measured atomic ratio (Pd : Mo : Te = 70.7 : 9.6 : 19.7) closely matches the 7 : 1 : 2 stoichiometry of $Pd_7MoTe_2$. **Figure 1G** shows high-resolution plan-view TEM images of $MoTe_2$ and the force-induced nanowire. The nanowire exhibits a distinct hexagonal lattice with an in-plane spacing of 0.22 nm, in agreement with the thermally grown $Pd_7MoTe_2$ structure (*17*). Moreover, high-resolution TEM imaging at the interface between force-induced $Pd_7MoTe_2$ and $MoTe_2$ (Fig. S5) reveals a sharp boundary and an ordered atomic structure of the adjacent $MoTe_2$, indicating that the force-induced growth process inflicts minimal damage and preserves the materials' structural integrity.

## B. The effect of temperature and force

To understand the novel mechanochemically induced 2D diffusion of Pd to form $Pd_7MoTe_2$, we systematically explored the effects of temperature and force on the growth rate. The growth rate was quantified by scanning the AFM probe along single lines across the $Pd_7MoTe_2$ and $MoTe_2$ regions at different applied forces and temperatures, and calculating the length of the grown $Pd_7MoTe_2$ nanowire per cycle based on the final nanowire geometry (see Methods). As shown in **Fig. 2A**, we measured the growth rate as a function of force from 40 °C to 140 °C. Mechanochemically induced growth of $Pd_7MoTe_2$ was achieved at all tested temperatures down to 40 °C, whereas without applied force the reaction requires temperatures near 200 °C (*15*). For temperatures below 120 °C, a critical load of 7.8 nN is required to initiate $Pd_7MoTe_2$ growth, increasing to 572 nN as the temperature decreases to 40 °C. At a fixed temperature, the normal load significantly influences the growth rate; for instance, the growth rate increases by a factor of seven as the normal load rises from 68 nN to 1150 nN at 140 °C, indicating that the growth of $Pd_7MoTe_2$ at the front might be a force- or stress-activated process (*20, 22-25, 30-33*). In the present experiments, the scanning speed of the AFM tip (~8 μm/s) is much higher than the net growth rate of the 2D $Pd_7MoTe_2$. No observable difference in the growth rate is observed between different scan speeds (Fig. S6). We note that while the instantaneous growth can be fast when stress is applied at the growth front (Fig. S7), the net growth rate remains slow because of the time intervals between successive passes. In addition to force, the influence of temperature can be seen clearly by plotting the logarithm of the growth rate against inverse temperature (**Fig. 2B**), showing a nearly linear decrease. In mechanochemical processes, these stress- and temperature-dependent growth rates Γ (nm of length/cycle in this case) is often well-described by a first-order stress-activated Arrhenius model attributable to Eyring (*25, 33*):

$$\Gamma = \Gamma_0 e^{-\left(\frac{\Delta G_{act}}{k_B T}\right)} \tag{1}$$

where $\Gamma_0$ is a prefactor related to an effective attempt frequency, $k_B T$ is the thermal energy, and $\Delta G_{act}$ the free energy of activation given by

$$\Delta G_{act} = \Delta U_{act} - \sigma \Delta V_{act} \tag{2}$$

where $\Delta U_{act}$ is the internal activation energy (the stress-free activation barrier), σ is the stress component that lowers $\Delta U_{act}$, and $\Delta V_{act}$ is the activation volume. $\Delta V_{act}$ signifies the sensitivity of $\Delta U_{act}$ to stress, and its magnitude is often attributed to the volume change occurring when reactants pass through their transition state and thus is an atomic-scale quantity (*31, 34, 35*). While this relation has been seen in many bulk mechanochemical experiments, nanoscale experiments exploring mechanochemical reactions are limited in number (*34*), with only a handful involving bond forming reactions (as opposed to wear, or pulling apart of molecules) (*20-22*).

Determining σ for a given applied tip-sample force is non-trivial, as the $MoTe_2$ and Pd are part of a nanoscale multi-material stack capped by multilayer hBN, all on a substrate. To estimate the stress, we calculated the maximum in-plane stress σ in the $MoTe_2$ based on FEM simulations of the strain distributions within the stack and $MoTe_2$'s known elastic constants (**Fig. 2C**, Methods, and Fig. S8). A global fit of the growth rate as a function of σ and temperature based on the stress-activated Arrhenius model (Eq. 1 and 2) fits the experimental results very well (**Fig. 2D**). The extracted activation parameters of $\Delta U_{act} = 0.34 \pm 0.01$ eV and $\Delta V_{act} = 6.5 \pm 0.6$ Å$^3$ are physically reasonable for atomic-level mechanochemical processes (*30*). Collectively, this strongly suggests that initial stages of $Pd_7MoTe_2$ growth at the front occur through a stress-assisted thermally activated process.

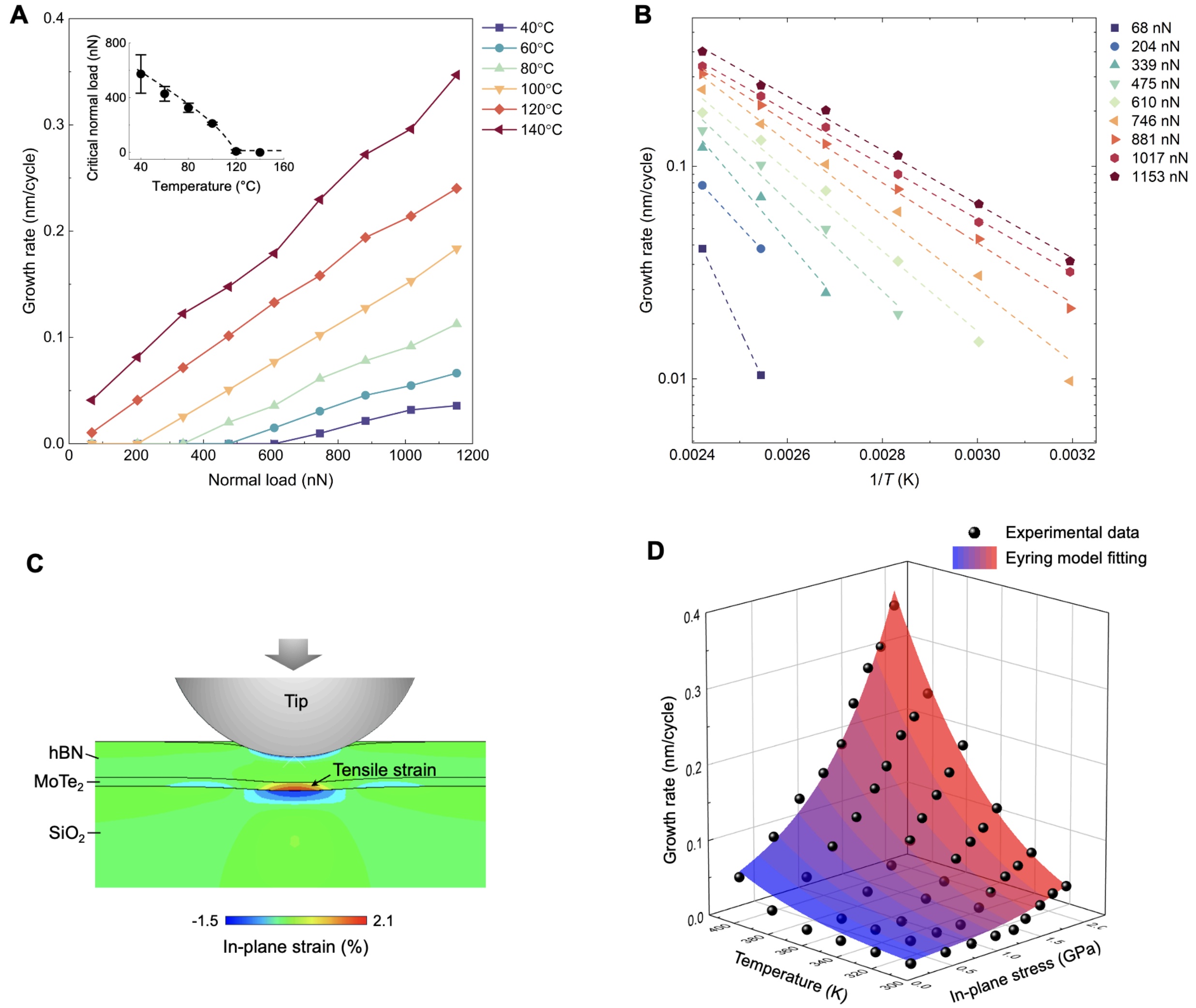

**FIG. 2**. **The effect of force and temperature on the growth rate of $Pd_7MoTe_2$.** (**A**) Growth rate versus normal load data measured under varying temperatures. Inset: critical normal load to initiate $Pd_7MoTe_2$ growth vs. temperature. (**B**) Logarithm of the growth rate versus inverse temperature for different normal loads. (**C**) FEM simulation shows the in-plane strain distribution with an applied normal load of 1153 nN. The maximum strain at each load is then used to calculate the maximum in-plane stress for (D). (**D**) Growth rate as a function of temperature and maximum in-plane stress in the $MoTe_2$, fitted using a stress-activated Eyring model (Eq. 1 and 2).

### C. *Ab initio* simulations of Pd growth on $MoTe_2$

To explore and understand the mechanochemical process by which $Pd_7MoTe_2$ forms at the growth front, *ab initio*-Grand Canonical Monte Carlo (ai-GCMC) calculations (*36-38*) were used to simulate the interaction of Pd with $MoTe_2$. We focus on modeling the nanoscale growth at the front, rather than the larger-scale transport of Pd, because the AFM experiments suggest that, remarkably, the main energy barrier for the growth is only at the front, where long-distance Pd transport via the preformed compound occurs spontaneously (Fig. S7). Freestanding monolayer $MoTe_2$ interacting with Pd was modeled for computational efficiency. The $SiO_2$ substrate and hBN layer are not included in the simulations; this is justifiable as hBN only interacts weakly with $MoTe_2$ via van der Waals forces, Pd likely bonds only weakly to $SiO_2$ (*39*), and thermally driven growth occurs on suspended samples with only Pd and the TMDs present (*15*).

**Fig. 3A** shows typical atomic configurations during the growth of Pd on the 1T′ $MoTe_2$ monolayer surface (1T′ $MoTe_2$ was used as the substrate since, as shown in Fig. S9, $Pd_nMoTe_2$ with $n > 0.2$ was significantly more stable energetically for 1T′ vs monolayer 2H $MoTe_2$). The first Pd atom (equivalent to a Pd stoichiometry of $n$=0.06 for the chosen simulation cell size) preferentially adsorbs at hollow sites above a central Mo atom and bonds to three Te atoms with bond lengths of ≈2.6 Å and a first-atom adsorption energy of 0.89 eV (unfavorable with respect to bulk Pd). Subsequent Pd atoms tend to adsorb near one another, forming a Pd-Pd-Pd chain or triangle to stabilize them. The Pd-Pd bond lengths vary from ≈2.8 Å to 2.98 Å, which are longer than in the bulk phase (2.784 Å). As $n$ increases, each Pd atom bonds to multiple other Pd atoms, and the second Pd layer starts to develop prior to the completion of the first layer. Flat layer-by-layer growth of Pd on the $MoTe_2$ surface would induce substantial strain due to the intrinsic lattice mismatch, and indeed flat growth is not observed in ai-GCMC. Instead, with atom addition, the Pd gradually changes from an initially amorphous structure to an FCC-like configuration, with the Pd <111> axis tilted by ≈22° with respect to the $MoTe_2$ <0001>, decreasing slightly as more Pd is added (Fig. S2). Simulated TEM images calculated from these ai-GCMC structures (*40*) (see Methods), *e.g.* for $n$=3.5 and 6.2, show a slightly distorted hexagonal pattern (**Fig. 3B**, top and middle panels respectively), consistent with the experimental TEM image in **Fig. 1G** (right image). The tilt angle suddenly reduces to ≈9° above $n$=7 (Fig. S2), and the corresponding simulated TEM image (**Fig. 3B**, bottom panel) no longer shows the clear hexagonal contrast seen in experiments; this is associated with defect formation as explained further below.

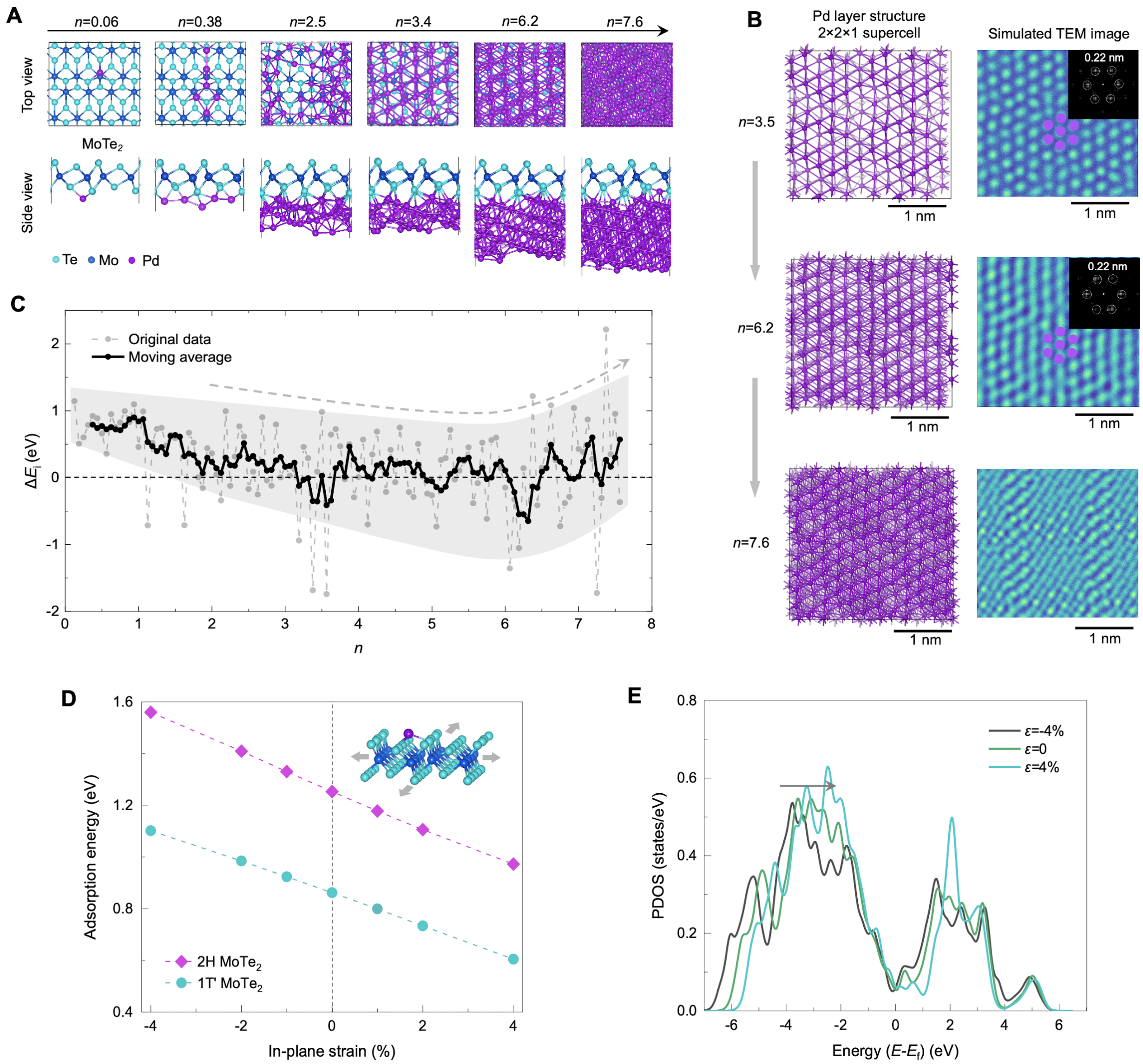

**FIG. 3. ab initio-GCMC simulation of 2D Pd growth on $MoTe_2$.** (**A**) Snapshots of representative configurations of Pd atom growth on 1T′-$MoTe_2$ from ai-GCMC, at different values of n for $Pd_nMoTe_2$. Blue, cyan, and magenta spheres represent Mo, Te, and Pd atoms respectively. (**B**) Typical atomic structures of the Pd atoms within $Pd_nMoTe_2$ with $n$=3.5, 6.2 and 7.6 (left column), and corresponding simulated TEM images (right column) and their Fourier transforms (insets). (**C**) The incremental formation energy ($\Delta E_i$) with respect to bulk Pd vs. $n$. Grey and black filled circles represent the ai-GCMC dataset and a five-point moving average respectively. (**D**) Adsorption energy of a single Pd atom with respect to bulk Pd as a function of in-plane strain of monolayer 2H-$MoTe_2$ and 1T′ $MoTe_2$, each composed of 4 formula units. The

% strain value is calculated as $\varepsilon=100\times(l-l_0)/l_0$, where $l$ and $l_0$ are the instantaneous lattice parameter and the optimized lattice parameter at zero strain respectively. Negative and positive values indicate compressive and tensile strain, respectively. (**E**) The partial density of states (PDOS) of Te(p)-orbitals vs. $\varepsilon$. The p-orbital energies approach the Fermi energy under tensile strain, suggesting a strong Te-Pd interaction, favoring Pd adsorption.

We further calculated the incremental (per atom) formation energy as a function of Pd stoichiometry $n$. As shown in **Fig. 3C**, the incremental formation energy is initially positive, indicating that Pd adsorption onto bare $MoTe_2$ is unfavorable, because Pd-Pd metallic bonding in the bulk is stronger than the Pd-Te interaction on the $MoTe_2$ surface. In experiments, either heat or mechanical force must be applied to provide additional energy to initiate $Pd_nMoTe_2$ growth, which is consistent with the observation of this energy barrier to initial Pd growth in the simulations. As more Pd atoms are added to the $Pd_nMoTe_2$, the incremental energy decreases, indicating Pd atoms increasingly prefer to move from bulk Pd (*e.g.*, like the Pd source in the experiments) to the $Pd_nMoTe_2$ surface. That is reasonable since as more Pd atoms migrate to the $MoTe_2$ surface, the formation of Pd-Pd metallic bonds and Te-Pd interface bonds leads to a reduction in incremental energy. However, the moving average of the incremental energy reaches a minimum near $n$=6 and thereafter increases, becoming nearly always positive by $n$=7.2. These energetic data support the growth of $Pd_nMoTe_2$ terminating around $n$=7.

The computationally observed growth proceeds in three stages. (i) Nucleation: In the early stages ($n$<2), Pd growth is unfavorable and thus requires an input of energy via heat or stress. The adsorbed Pd is mostly amorphous, due to the small number of adsorbed Pd atoms lacking sufficient neighbors to become FCC-like. (ii) Growth: Once $n$>2 is reached, the Pd rearranges to become increasingly FCC, with tilted <111> axis relative to the $MoTe_2$ <0001> by ≈22°. Calculations of bond length variations with Pd stoichiometry (Fig. S10) show changes for $n$ > 3, indicating that the inclined Pd surface minimizes lattice mismatch, achieving minimal strain energy. Our analysis suggests that this structure (for 2<$n$<7) is a minimum-energy compromise between the Pd/$MoTe_2$ interfacial energy, the Pd film energy including elastic and defect energetics, and the Pd surface energy. (iii) Transformation: If we force the system to go above n>7, we find computationally a transformation to a structure with a much smaller tilt ≈9° and dislocations in the crystalline structure. The common nearest-neighbor analysis also indicates that $n$=7.6 structure transformed to a bulk FCC structure with dislocations (**Fig 3B** and Fig. S11). This is evidence of an energetic crossover to another structure with lower interfacial energy, surface energy, and strain energy at the cost of higher bulk defect energy. After this point, the simulations predict that additional Pd atoms do not prefer to join the $Pd_7MoTe_2$ film and growth ceases. In the experiment, the transformation is perhaps kinetically suppressed, as this "transformation" structure is not observed; instead, lateral growth to extend the $Pd_7MoTe_2$ is preferred. Thus, a balance of interfacial, surface, and bulk energetics leads to self-limiting growth of $Pd_7MoTe_2$.

As demonstrated by the FEM simulation, the local compressive stress applied by the AFM tip leads to in-plane tensile strain of the $MoTe_2$. To model this mechanical effect, the adsorption energy of single Pd atom on $MoTe_2$ was calculated *vs.* in plane strain (**Fig. 3D**). Pd adsorption is stabilized by increasing tensile strain, while compressive strain exhibits the opposite trend. This indicates that tensile strain facilitates Pd adsorption by weakening Mo-Te bonds, thereby promoting the formation of stronger Pd-Te interactions, as evidenced by the shift in the *p*-state peaks in the partial density of states (PDOS) (**Fig. 3E**). It may also promote better registry between the Pd and the $MoTe_2$.

**D. Superconductivity in mechanochemically written $Pd_7MoTe_2$ nanostructures**

To demonstrate the exceptional precision and quality of the mechanochemical nano-writing method in fabricating finely designed structures, and to verify that the mechanochemically induced material exhibits superconductivity like its thermally grown analog, we used the nano-writing method to directly construct a Hall bar structure of $Pd_7MoTe_2$, again with hBN encapsulation (**Fig. 4A**). The evolution of the Hall bar structure during the writing process is captured through a series of AFM topographic images (**Fig. 4B**) where a $Pd_7MoTe_2$ nano-bar of ~3.8 μm in length and ~200 nm in width is fabricated. The nano-bar is subsequently connected with side contacts, also written by AFM, for four-probe measurement. **Fig. 4C** plots the four-probe longitudinal resistance ($R_{xx}$) of the $Pd_7MoTe_2$ Hall bar (Sample D2) *vs.* temperature ($T$), with a measurement configuration indicated in the inset of **Fig. 4C**. The as-written $Pd_7MoTe_2$ exhibits a characteristic superconducting transition near a critical temperature ($T_c$) of ≈ 0.8 K, below which zero resistance is observed (**Fig. 4C**). **Figure 4D** displays the characteristic nonlinear $I$-$V$ curves taken at various $T$, plotted together with the differential resistance ($dV/dI$) at base temperature, which demonstrate clearly the sharp critical current behavior. We note that the DC current ($I$) is swept from positive to negative values in our measurement, and the observed asymmetry between the positive and negative critical currents may be attributed to heating effect before entering the zero-resistance state on the positive side. This is a known mechanism for producing such asymmetric appearance in critical currents. The magnetic field ($B$, applied normal to the 2D plane) dependence of the critical current (**Fig. 4E**) shows that superconductivity is suppressed above 1.2 T. All the superconducting characteristics observed here in the mechanochemically written $Pd_7MoTe_2$ (**Fig. 4C-4F**) are consistent with the previous studies (*15-17*) on thermally grown $Pd_7MoTe_2$. This confirms not only the material's identity but also the high crystal quality achieved through the force-induced growth mechanism.

Structures with even smaller sizes than the example above can be fabricated with this method. As shown in Fig. S12, we fabricated superconducting nanowires with <50 nm, nanowire gap junctions with gaps <50 nm, fence-like structures with sharp corners, and a superconducting quantum interference device (SQUID)-like structure with <50 nm gaps at the junctions.

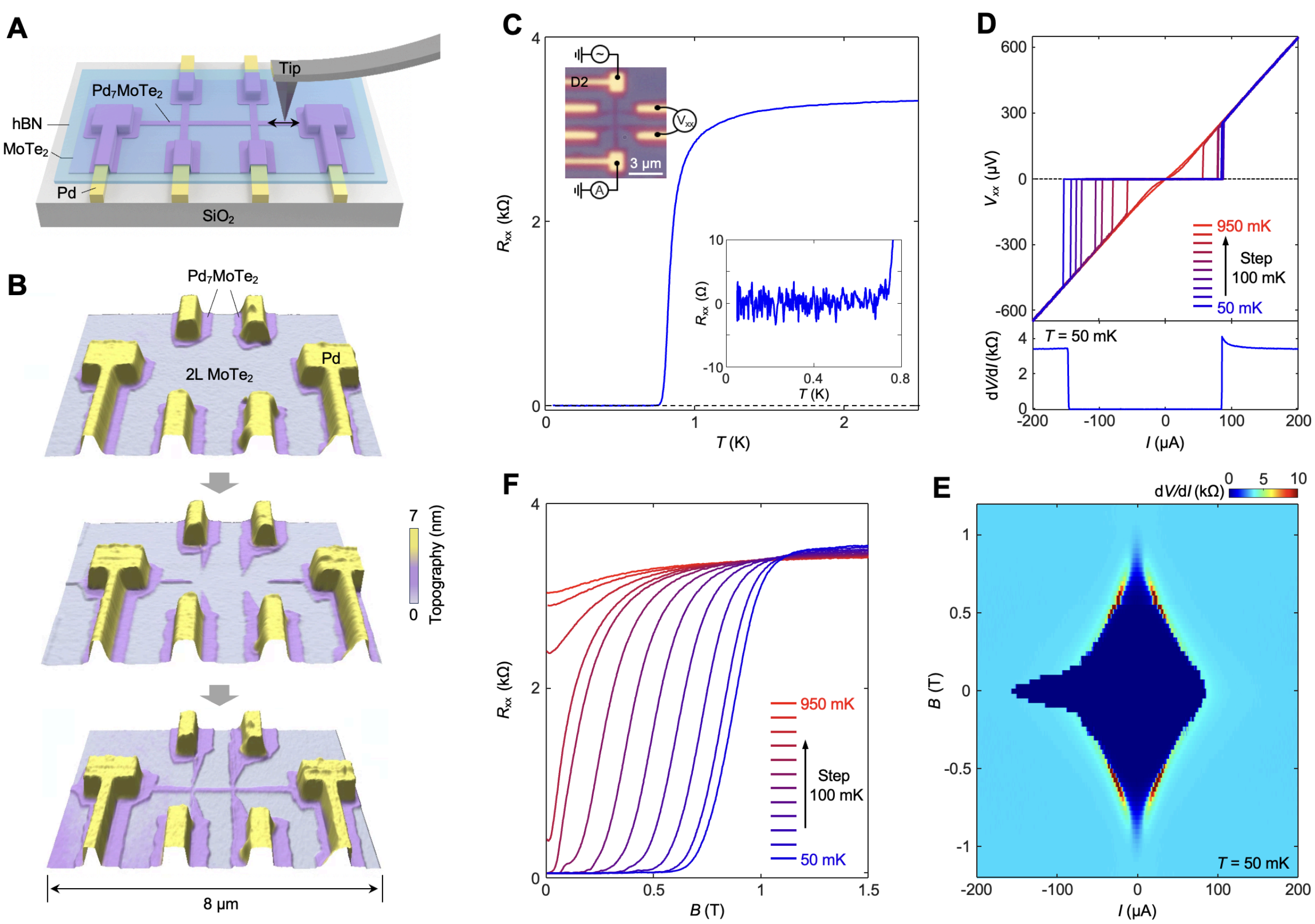

**FIG. 4. Encapsulated nano-writing of 2D superconducting $Pd_7MoTe_2$.** (**A**) Schematic illustration of the experimental setup used for writing the $Pd_7MoTe_2$ Hall bar structure. (**B**) 3D

AFM topographic images illustrating the evolution of the $Pd_7MoTe_2$ Hall bar structure at the initial, intermediate, and final stages of growth. (**C**) Longitudinal resistance $R_{xx}$ versus temperature for the $Pd_7MoTe_2$ nanowire. The measurement configuration is shown in the top-left inset, and the bottom-right inset provides a magnified view of $R_{xx}$ below 0.8 K, confirming the zero resistance. (**D**) Nonlinear I-V curves measured at various T (top) and differential resistance dV/dI as a function of applied DC current I at 50 mK (bottom). (**E**) A dV/dI map as a function of applied DC current I and perpendicular magnetic field B at the temperature of 50 mK. (**F**) The resistance of $R_{xx}$ as a function of B measured at various temperatures.

## III. DISCUSSION

Using mechanical force, we drive precisely controlled formation of atomically thin Pd confined in a 2D space, that subsequently reacts with the original $MoTe_2$ above it to form crystalline $Pd_7MoTe_2$, with controlled lateral dimensions down to 50 nm. Force accelerates the reaction kinetics exponentially, remarkably reducing the temperature needed to form $Pd_7MoTe_2$ from 200°C to 40°C, with the mechanochemically formed $Pd_7MoTe_2$ exhibiting superconductivity consistent with thermally formed $Pd_7MoTe_2$. DFT elucidates the stages of Pd growth, rationalizing the self-limiting 7-atom-layer growth and showing that $Pd_7MoTe_2$ formation is initially accelerated mechanically due to tensile strain induced in the $MoTe_2$, consistent with experiments and FEM simulations. Key benefits derived from mechanochemical nano-writing include precise dimensional control, a substantially reduced need for heating (valuable for devices where thermally sensitive materials are present), and *in situ* real time monitoring and characterization of the synthesized product, all demonstrated by the fabrication of multiple nanoscale device structures at temperatures well below the strain-free formation temperature of $Pd_7MoTe_2$. Due to the generally-applicable manner in which mechanical stress can drive reactions over energy barriers, this approach can be applied to form other metal-TMD compounds and to other TMD moiré materials (*17, 41-45*). This highlights the method's potential for on-chip integration of next-generation nanoelectronic and quantum devices, including offering an appealing route to engineering superconductivity at the nanoscale in these rich physical systems.

## ACKNOWLEDGMENTS

Work by Wu lab is mainly supported by AFOSR through awards FA9550-23-1-0140 and FA9550-25-1-0354 to S.W. Authors S.Z., A.S., Y.B., T.L., C.Q., A.M., A.M.R., and R.W.C. acknowledge support from the NSF Center for the Mechanical Control of Chemistry under grant #CHE-2303044. S.W. and L.M.S. and N.Y. acknowledge the support by NSF through the Materials Research Science and Engineering Center (MRSEC) program of the National Science Foundation (DMR-2011750) and its partial support to Princeton's Imaging and Analysis Center (IAC). S.W. acknowledges the support from Princeton Catalysis Initiative and the Gordon and Betty Moore Foundation through Grant GBMF11946. S.Z. acknowledge the support by the Fundamental Research Funds for the Central Universities (226-2025-00104). K.W. and T.T. acknowledge support from the JSPS KAKENHI (Grant Numbers 21H05233 and 23H02052) and World Premier International Research Center Initiative (WPI), MEXT, Japan. The AFM work was carried out at the Singh Center for Nanotechnology, which is supported by the NSF National Nanotechnology Coordinated Infrastructure Program under grant NNCI-2025608. We thank Q. Wang and Q. Li for discussions and help with FEM calculation. Computational support was provided by the National Energy Research Scientific Computing Center (NERSC), a U.S. Department of Energy, Office of Science User Facility located at Lawrence Berkeley National Laboratory, operated under Contract No. DE-AC02-05CH11231.